\documentclass[a4paper,11pt]{article}
\usepackage{jcappub} 
\usepackage{lineno}

\usepackage{lmodern}
\usepackage{mathtools}
\usepackage{enumitem}
\usepackage{verbatim}
\usepackage{multirow}
\usepackage{graphicx}
\usepackage{physics}
\usepackage{xspace}

\newcommand{\ChargeConstraintMonePN}{\ensuremath{3 \times 10^{-34}}}
\newcommand{\ChargeConstraintAH}{\ensuremath{2 \times 10^{-26}}}

\title{\boldmath Bounds on the charge of the graviton using gravitational wave observations.}

\author[a]{S. Nair,}
\author[b]{A. Vijaykumar,}
\author[a,c]{S. Sarkar}

\affiliation[a]{Indian Institute of Technology Gandhinagar,\\Gandhinagar, Gujarat 382055, India}
\affiliation[b]{Canadian Institute for Theoretical Astrophysics,\\University of Toronto, 60 St. George Street, Toronto, ON M5S 3H8, Canada}
\affiliation[c]{Instituto Galego de F´ısica de Altas Enerx´ıas
(IGFAE),\\ Universidade de Santiago de Compostela, Spain.}

\emailAdd{sreejithnair@iitgn.ac.in}
\emailAdd{aditya@utoronto.ca}
\emailAdd{sudiptas@iitgn.ac.in }

\abstract{If the graviton possesses a non-zero charge $q_g$, gravitational waves (GW) originating from astrophysical sources would experience an additional time delay due to intergalactic magnetic fields. This would result in a modification of the phase evolution of the observed GW signal similar to the effect induced by a massive graviton. As a result, we can reinterpret the most recent upper limits on the graviton's mass as constraints on the joint mass-charge parameter space, finding $\abs{q_g}/{e} < \ChargeConstraintMonePN{} $ where $e$ represents the charge of an electron. Additionally, we illustrate that a charged graviton would introduce a constant phase difference in the gravitational waves detected by two spatially separated GW detectors due to the Aharonov-Bohm effect. Using the non-observation of such a phase difference for the GW event GW190814, we establish a mass-independent constraint $|q_g|/e < \ChargeConstraintAH{}$. To the best of our knowledge, our results constitute the first-ever bounds on the charge of the graviton. We also discuss various caveats involved in our measurements and prospects for strengthening these bounds with future GW observations.\\}

\begin{document}
\maketitle
\flushbottom

    \section{Introduction} 
   In general relativity (GR), gravitational waves (GWs) travel at the speed of light, asserting that gravitons must have zero rest mass.  
However, in search of new physics, we must test this key idea and look for potential deviations.
    The detection of GWs from merging compact binary systems~\cite{LIGOScientific:2016aoc, LIGOScientific:2018mvr, LIGOScientific:2020ibl, KAGRA:2021vkt} by the LIGO-Virgo-KAGRA (LVK) collaboration~\cite{LIGOScientific:2014pky, VIRGO:2014yos, KAGRA:2020tym} has opened up a new avenue for testing such essential foundations of Einstein's theory of general relativity \cite{LIGOScientific:2016lio,LIGOScientific:2018dkp, LIGOScientific:2019fpa, LIGOScientific:2020tif,LIGOScientific:2021sio}. 
    
    To investigate any potential deviations from GR, one could consider specific alternative models or theories and constrain their parameters from observations using the gravitational waveform calculated within the theory. However, calculating waveforms in alternative theories of gravity has proved to be difficult.
    But suppose we consider GWs generated in a modified theory of gravity that lacks a mathematical framework to comprehend the GW generation process. Nevertheless, since the GWs propagate over cosmological length scales, cumulative corrections affecting the GW propagation could become dominant over other effects related to its generation. Such propagation effects offer an excellent possibility for theory-agnostic tests of new gravitational physics and allow for stringent bounds on parameters quantifying the violation of general relativity. A specific illustration of such an approach involves constraining the mass of the graviton through the GW observations, as presented in \cite{Will:1997bb, Mirshekari:2011yq, LIGOScientific:2016lio, LIGOScientific:2020tif, LIGOScientific:2021sio}. 
    To constrain the graviton mass, one computes the time difference of detection, $\Delta t_o$, of two gravitons emitted at a time difference $\Delta t_e$ apart at the source with an additional mass-dependent term in the dispersion relation of the GWs \cite{Will:1997bb}. This non-zero mass adds an additional contribution to the GW phasing at $-1$ post-Newtonian (PN) order. Using this modified phasing formula, the latest GW observations bound the mass of the graviton ($m_g$) to be $m_g\leq 1.27\times 10^{-23}\,{\rm eV}/c^2$ \cite{LIGOScientific:2021sio}.
    
    In a similar phenomenological spirit, let us ask whether the graviton can be endowed with a non-zero electric charge. Here, we note earlier attempts at constraining the electric charge associated with neutrinos and photons \cite{Barbiellini:1987zz, Raffelt:1994rb, Altschul:2007xf, Altschul:2007vca} using astrophysical observations. These observations utilized the interaction of the particle with the ambient magnetic field and the resulting additional time delay for their bounds. Interestingly, the relative time delay induced by the mass and the electric charge of a particle of energy $E$ has the same $E^{-2}$ scaling \cite{Raffelt:1994rb, Barbiellini:1987zz, Cocconi:1988xi}. This suggests that if we consider a massive graviton, the measured mass may be indistinguishable from the possible electric charge of the graviton.
     
    It should be noted that attributing an electric $U(1)$ charge to a massless particle is not straightforward.
     The electric charge of the graviton will be associated with a coupling of the gravitational field to the background electromagnetic field via an $U(1)$ interaction. Attributing such a charge to a massless graviton may lead to inconsistencies. For instance, attempts at constructing quantum theories of higher spin fields coupled to the $U(1)$ gauge field resulted in inconsistencies in the quantization of such theories if the fields are massless \cite{Berends:1979rv, deWit:1979sib, Johnson:1960vt}. Additionally, the Weinberg-Witten theorem argues that in a higher spin ($j > 1/2$; $j=2$ for graviton) theory that allows the construction of a Lorentz-covariant conserved four current, one cannot have a charged massless particle \cite{Weinberg:1980kq}. However, there are no apparent field theoretic obstructions on a massive graviton like the one considered in \cite{Will:1997bb, LIGOScientific:2016lio, LIGOScientific:2020tif, LIGOScientific:2021sio} having a $U(1)$ charge.
   
    The above observation, in conjunction with the possible degeneracy of the effect of putative mass with the electric charge of the graviton, makes the phenomenology of the graviton charge particularly relevant. 
    In this work, we shall consider the effects of an electrically charged massive graviton on the GW phasing and impose simultaneous bounds on its charge and mass. We shall compute the energy-dependent time delay induced on the propagation of a charged massive graviton due to its interaction with the intergalactic magnetic field in the same spirit as the effect of a non-zero charge on the propagation of the neutrino \cite{Barbiellini:1987zz} or the photon \cite{Raffelt:1994rb, Altschul:2007xf}. This time delay can be associated with the observed GW phasing through the methods developed in \cite{Will:1997bb}; here, we explicitly demonstrate the degeneracy of the graviton charge and mass. Following this, we use the LVK observations to impose simultaneous constraints on the graviton's mass and charge. Our analysis results in an upper bound on the graviton charge in terms of the electron charge $e$ of $|q_g|/e< \ChargeConstraintMonePN{}$. Furthermore, based on previous work~\cite{Altschul:2007xf, Altschul:2007vca}, we show that, in the presence of a charged graviton, GWs detected at two detectors separated by a finite distance would have a phase difference; this phase difference can be understood as the Aharonov-Bohm phase~\cite{Aharonov:1959fk} due to the interaction of the charged particle with the intergalactic magnetic field. Using the measurement of the phase difference (consistent with zero) between the LIGO Hanford and LIGO Livingston detectors from GW190814, we place an independent bound of $|q_g|/e< \ChargeConstraintAH{}$ on the charge of the graviton. 
    \\
    
    \section{Imprint of charge on the waveform}
    \label{sec:imprint-on-waveform}
    
    Let us consider compact binaries with component masses $m_1$ and $m_2$, which radiate away energy through GWs, where we assume the radius of their circular orbit to vary adiabatically. This problem is well understood perturbatively within GR using the framework of restricted post-Newtonian (PN) formalism, which expresses the amplitude $A(f)=\mathcal{A}\, f^{-7/6}$ through the quadrupole approximation and the phase $\psi(f)$ is given by an expansion in powers of ${\beta(f) := v(f)/c}$ \cite{Blanchet:2001aw, Blanchet:2013haa}: 
    \begin{equation} 
    \label{waveform}
        h(f) = \mathcal{A}\, f^{-7/6}\, e^{i\, \psi(f)}\, ,
    \end{equation}
    where $f$ is the frequency of the emitted GW and ${v = (\pi G M f / c^3)^{1/3}}$ is the orbital speed of the binary with total mass $M = m_1 + m_2$.\\

    We shall follow the prescription outlined in Ref.~\cite{Will:1997bb} for our discussion. We will consider gravitational radiation emitted from such a binary with a frequency $f_e$ and the graviton associated with this radiation. If the graviton has a non-zero charge, its trajectory will be curved when exposed to a transverse magnetic field. This alteration of the trajectory will affect the time the graviton takes to reach the detector once emitted. This delay will be in addition to the delay caused by a non-zero mass, considered in \cite{Will:1997bb}.\\

    To compute the additional time delay caused by an alteration of the trajectory, we will perform an analysis similar to the neutrino charge estimation \cite{Barbiellini:1987zz}. Since the correlation length associated with the intergalactic magnetic field (IGM) is much shorter than the distance of propagation of the graviton, we need to account for the presence of multiple magnetic field domains along the path of propagation \cite{Altschul:2007xf,Altschul:2007vca}.\\
    
    \begin{figure}
    \label{bending picture}
        \centering
        \includegraphics[width=0.4\textwidth]{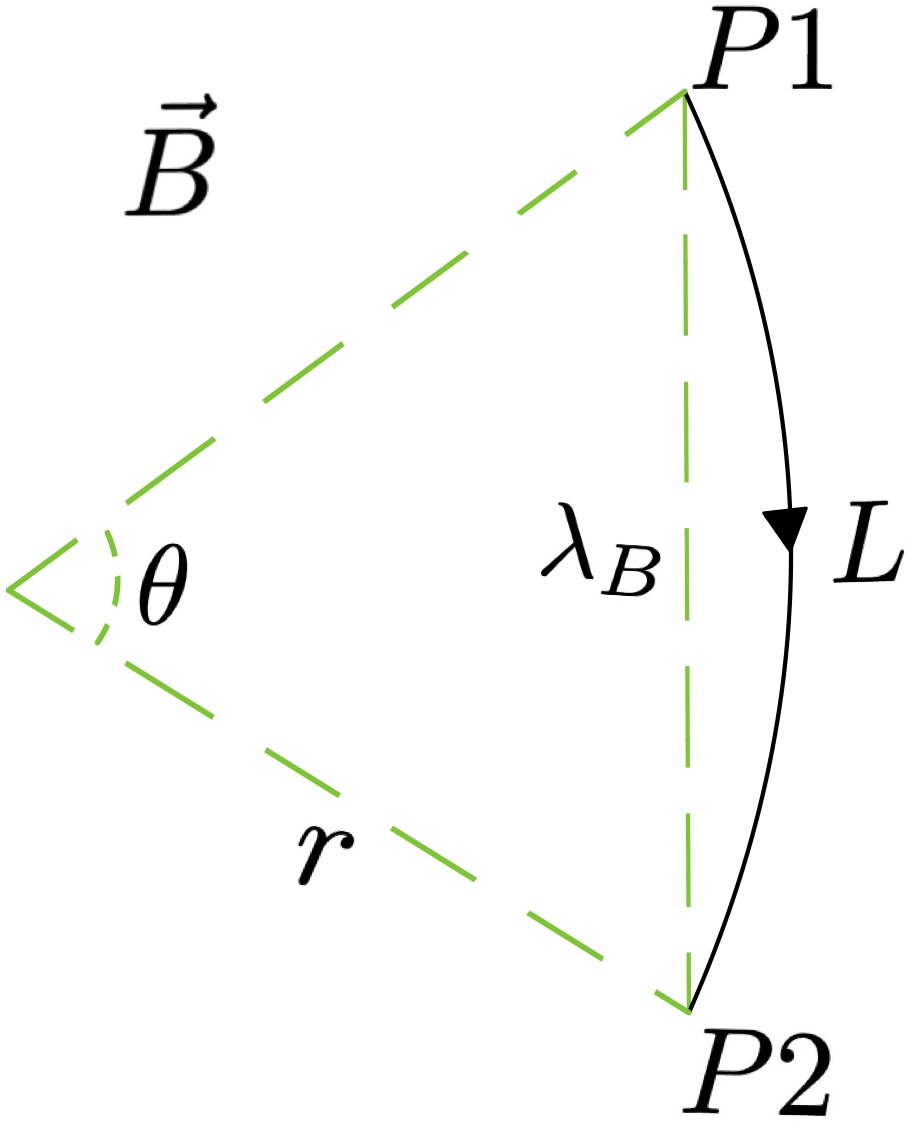}
        \caption{A schematic diagram illustrating the alteration of the trajectory as a charged graviton propagates from point $P1$ to $P2$ across a magnetic field domain of length $\lambda_B$. The transverse magnetic field ($\Vec{B}$) will cause the trajectory of the graviton to be curved, with a radius of curvature given by $r=\frac{p}{B q_g}$, $p$ being the momentum and $q_g$ being the charge of the graviton. The radii subtends an angle $\theta$, such that the distance travelled by the graviton is $L=r\theta$.}
    \end{figure}
    
    To start,  we note that if the intergalactic medium has a magnetic field of magnitude $B$ and correlation length  $\lambda_B$, the region between the source and the observer can be considered to be divided into $N$ magnetic field domains, each of length $\lambda_B$ such that the luminosity distance, $D_L=N\lambda_B$. A magnetic field of magnitude $(B)$ transverse to the direction of propagation will result in the trajectory of a charged graviton being curved, with a radius of curvature given by $r=\frac{p}{B q_g}$, $p$ being the momentum and $q_g$ being the possible charge of the graviton. This has been illustrated in Fig.~(\ref{bending picture}). The additional time in relation to the uncharged case that a charged massive graviton would take due to the presence of a transverse magnetic field to travel the distance $\lambda_B$ can now be calculated.\\
    
    By referring the schematic  in Fig.~(\ref{bending picture}), the additional time a graviton would take to traverse the curved path from $P1$ to $P2$ of length $L=r\theta$, $\theta$ being the angle subtended by the radii is given to be of the following form when $\theta$ is small \cite{Barbiellini:1987zz}\\ 
    \begin{equation}
    \label{bendcorr}
        \begin{aligned}
            \delta t_\lambda&=\frac{(L-\lambda_B)}{v_g}\\
            &\approx\frac{\lambda_B^2}{4 ! \, r^2}\Delta t_\lambda~.
        \end{aligned}
    \end{equation}
    In the above expression, $v_g$ is the magnitude of the three velocity of the massive graviton, $r=\frac{p}{B q_g}$ with $p$ being the momentum and $q_g$ being the possible charge of the graviton and $\Delta t_\lambda$ is the time a massive uncharged graviton of momentum $p$ would take to traverse the distance $\lambda_B$ and
    \begin{equation}
    \label{stprop}
        \begin{aligned}
            \Delta t_\lambda&=\frac{\lambda_B}{v_g} =\frac{\lambda_B E_g}{p c^2}\\
        \end{aligned}
    \end{equation}
    $E_g =h f_e $ is the energy of the graviton with frequency $f_e$. Using the above expression for $\Delta t_\lambda$, Eq.~\eqref{bendcorr} can be written as
    \begin{equation}
    \label{bendcorr2}
        \begin{aligned}
            \delta t_\lambda&=  \frac{\lambda_B^3 }{4 !\, c r^2 \sqrt{1-\frac{m_g^2c^4}{(hf_e)^2}}}= \frac{c\, \lambda_B^3 B^2 q_g^2}{4 !\,h^2 f_e^2\left(1-\frac{m_g^2c^4}{h^2 f_e^2}\right)^{\frac{3}{2}}},
        \end{aligned}
    \end{equation}
    where we have used the relativistic dispersion relation to express the momentum as a function of the mass and the energy of the graviton.\\
    
    Since there are $N$ such magnetic field domains between the source and the observer, the total additional time a charged massive graviton of frequency $f_e$ would take in relation to a chargeless massive graviton of the same frequency will be 
    \begin{equation}
    \label{tqtotal}
        \delta t_q\approx\sum_N p_\lambda \delta t_\lambda ~.
    \end{equation}
    Where $p_\lambda$ is the probability for the magnetic field inside the magnetic field domain to be aligned transverse to the direction of propagation, and for randomly oriented magnetic field domains this can be approximated as the probability for it to be aligned along four of the six cardinal directions, so $p_\lambda\approx\frac{2}{3}$. Now, we can use Eq.~\eqref{bendcorr2} to write Eq.~\eqref{tqtotal} as
    \begin{equation}
    \label{tqtotal2}
        \begin{aligned}
            \delta t_q&\approx\frac{2N}{3}\frac{c \lambda_B^3 B^2 q_g^2}{4 !h^2 f_e^2(1-\frac{m_g^2c^4}{h^2 f_e^2})^{\frac{3}{2}}}
            \approx \frac{2}{3} \frac{c\, D_L \lambda_B^2 B^2 q_g^2}{4! \, h^2 f_e^2\left(1-\frac{m_g^2c^4}{h^2 f_e^2}\right)^{\frac{3}{2}}}~.  
        \end{aligned}
    \end{equation}
    Having computed the additional time delay\footnote{We note that there might be an $\mathcal{O}(1)$ factor correction to the above expression that arises as a consequence of the assumptions made, like the IGM being of a fixed magnitude and suddenly changing direction at the edge of each magnetic field domain and equating $p_\lambda$ to $2/3$ \cite{Altschul:2007vca, Altschul:2007xf}. However, such corrections will not change the order of magnitude bound on the graviton charge.} a charged massive graviton would encounter while traversing a distance $D_L$, we will next compute the phase correction, which is expected at a GW detector \cite{Will:1997bb,Ghosh:2023xes}.\\
    
    If we consider two successive gravitons of frequencies $f_e$ and $f_{e'}$ emitted a time $\Delta t_e$ apart from a source at low redshift; they will reach the detector at luminosity distance $D_L$, $\Delta t_o$ apart in time, such that
    \begin{equation}
    \label{deltato}
        \begin{aligned}
            \Delta t_o&\approx\Delta t_e+\delta t^m_{f_e,f_{e'}} + \delta t^q_{f_e,f_{e'}}~.
        \end{aligned}
    \end{equation} 
    Where, $\delta t^m_{f_e,f_{e'}}$, is the expected time difference between two gravitons emitted at orbital frequencies $f_e$ and $f_{e'}$, as a result of the graviton being massive without any charge \cite{Will:1997bb}. 
    \begin{equation}
    \label{tm}
        \begin{aligned}
            \delta t^m_{f_e,f_{e'}}&\approx \frac{D_Lc^3m_g^2}{2h^2}\left(\frac{1}{f_e^2}\right)-\frac{D_Lc^3m_g^2}{2h^2}\left(\frac{1}{f_{e'}^2}\right).
        \end{aligned}
    \end{equation} 
    In the above expression, we used the assumption $m_gc^2<< hf_e$. While $\delta t^q_{f_e,f_{e'}}$, is the correction due to the possible change in the path of a charged and massive graviton. Using Eq.~\eqref{tqtotal2}, we can compute $\delta t^q_{f_e,f_{e'}}$ to be
    \begin{equation}
    \label{tq}
        \begin{aligned}
            \delta t^q _{f_e,f_{e'}}&\approx \frac{2}{3}\frac{c D_L \lambda_B^2 B^2 q_g^2}{4 !h^2 }\left(\frac{1}{f_e^2}\right)-\frac{2}{3}\frac{c D_L \lambda_B^2 B^2 q_g^2}{4 !h^2 }\left(\frac{1}{f_{e'}^2}\right)~.
        \end{aligned}
    \end{equation} 
    After assuming that the evolution of the binary is driven by gravitational back-reaction such that the associate $df_e/dt_e$ is well approximated by general relativity \cite{Will:1997bb, Ghosh:2023xes} and any correction to the generation of gravitational radiation due to the modified theory of gravity is negligible in relation to the terms proportional to the large luminosity distance $D_L$ we will get the following equation (see Appendix \ref{Appendix:Phase at detector.}) for the GW phase at the detector.
    \begin{equation}
    \label{phsemod}
        \begin{aligned}
            \psi(f)= & 2 \pi f \tilde{t}_c-\tilde{\Phi}_c-\frac{\pi}{4} + \sum_n\alpha^n f^\frac{2n}{3}-\left\{\frac{\pi  m_g^2 c^3}{h^2}+\left(\frac{2}{3}B^2\lambda_B^2\right)\frac{2 \pi q_g^2c }{4 ! h^2}\right\}\frac{D_L}{f},\\
        \end{aligned}
    \end{equation}      
    where all integration constants have been absorbed into $\Tilde{t}_c$ and $\Tilde{\phi}_c$. $\alpha^n$ are the usual PN terms from GR.\\

    Comparing the GW phasing given in Eq.~\eqref{phsemod} for a charged massive graviton with that of a massive graviton in Eqn. 3.8 of \cite{Will:1997bb} suggests that if we entertain the possibility of $U(1)$ charge on the graviton, the estimates on the mass of the graviton $m_g$ can be interpreted as a bound on the effective mass, $m_{\rm eff}$ of a charged massive graviton given by\\ 
    \begin{equation}
    \label{effm}
        m_{\rm eff}=\sqrt{  m_g^2+\left(\frac{2}{3}B^2\lambda_B^2\right)\frac{2  q_g^2 }{4 ! c^2}}~.
    \end{equation}
    One may argue strong magnetic fields can exist in the vicinity of a binary merger and those need to be taken into account while calculating the GW phasing in the presence of a charged graviton. However, even if these magnetic fields are strong they would not be expected to stay coherent over long length scales like the intergalactic magnetic field, thus minimizing their effect on the GW phasing. Hence, we will assume that the $B$ and $\lambda_B$ that enters Eq.~\eqref{phsemod} are the same as their values in the intergalactic magnetic field.

    \section{Detector dependent phase shifts due to graviton propagation}
    \label{sec:imprint-phase-difference}

    In addition to the above de-phasing of the GW resulting from a non-zero graviton charge, we can also expect a relative frequency-independent phase shift between GW signals from the same source observed by two separate detectors. This is a consequence of the well-known Aharonov-Bohm effect~\cite{Aharonov:1959fk}. A similar analysis has been carried out to bound the charge of a photon \cite{Altschul:2007xf,Altschul:2007vca}; here we extend the analysis to bound the graviton charge.\\
    
\begin{figure}
    \label{Diagram}
        \centering    
        \includegraphics[width=0.43\textwidth]{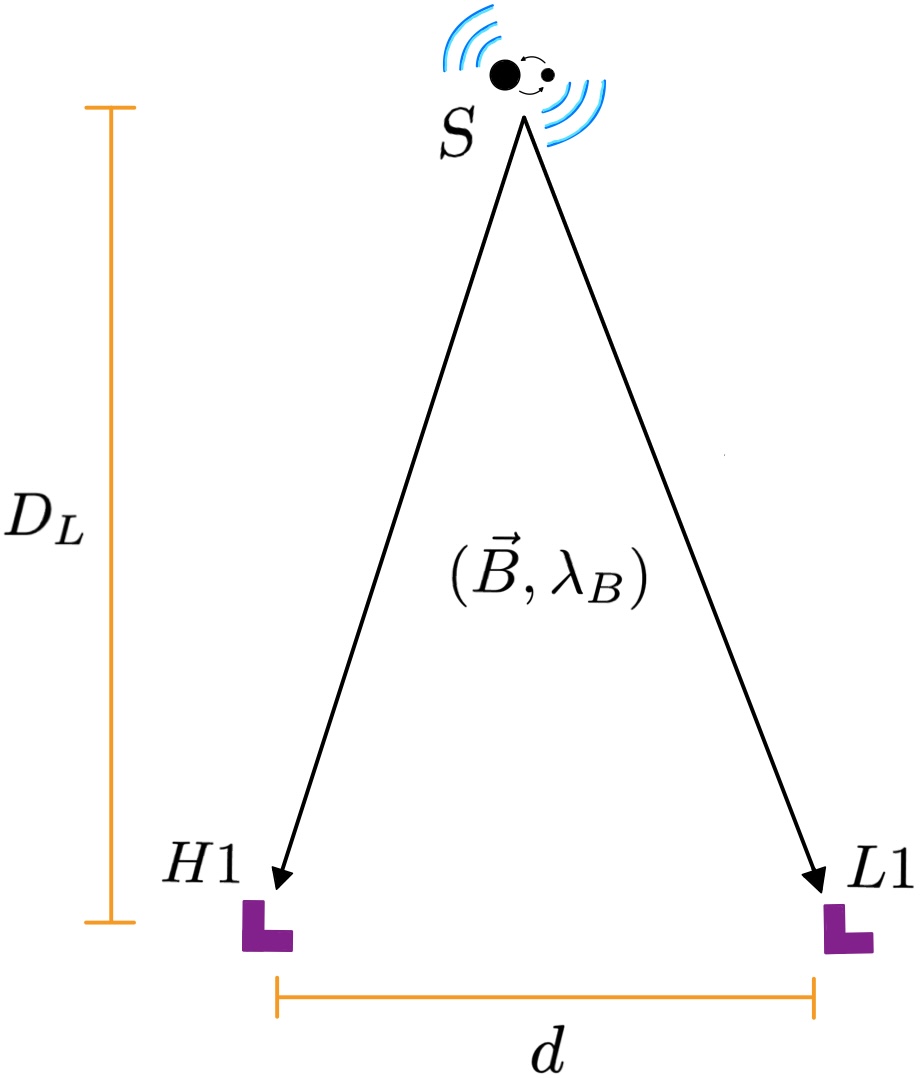}
        \caption{A schematic diagram illustrating the propagation of a graviton from the source $(S)$ to the two detectors $H1$ and $L1$ separated by a distance $d$, after traversing a luminosity distance of $D_L$. We assume that the intergalactic space between the GW source and the detector has a randomly oriented magnetic field of magnitude $B$ and correlation length $\lambda_B$. The magnetic field contributes to the total flux in the enclosed triangular region only when pointed in or out of the plain of the paper, behaving like a random walk of step size $D_L/3\lambda_B$.}
    \end{figure}

    
    Let us assume that the coupling of the electromagnetic field to the graviton follows an effective Lagrangian of the form $L_I=-\frac{q_g}{c} v^\mu A_\mu$, where $q_g$ is the charge of the graviton, $A^\mu$ is the $U(1)$ connection and $v^\mu$ is the 4-velocity of the graviton. The quantum phase difference between the two gravitons detected at separate locations can be expressed as
    \begin{equation}
    \label{Aphase}
    \Delta\phi=\frac{q_g\Phi}{\hbar c}~.
    \end{equation}      
    Where $\Phi$ is the total flux in the region enclosed by the two trajectories. For randomly oriented magnetic fields with a typical magnitude $B$ and correlation length $\lambda_B$, we have $D_L/\lambda_B$ magnetic field domains between the source located at a distance $D_L$ away and the observer.
    Since the magnetic field domains are randomly oriented, only one-third of all the domains will contribute to the phase, behaving like a random walk. We can compute the mean distance from the origin for this random walk after $D_L/3\lambda_B$ steps to be $\sqrt{2D_L/3\pi\lambda_B}$. Then, the total flux in the triangular region as illustrated in Fig.~(\ref{Diagram}) reads \cite{Altschul:2007xf,Altschul:2007vca}
    \begin{equation}
    \label{eq:aharonov-bohm-constraint}
    \Phi = \sqrt{\frac{D_L\lambda_B}{6\pi}}Bd~,~q_g = \dfrac{\sqrt{6 \pi}\hbar c}{\sqrt{\lambda_B} B} \left(\dfrac{\Delta \phi}{\sqrt{D_L} d}\right)
    \end{equation}      
    Here, $d$ is the separation between two detectors. The phase difference derived above is associated with distinct but coherent gravitons generated at the source observed at two different detectors. Nevertheless, similar to the case of photons, we can equate the classical GW phase difference observed and quantum phases of a graviton through the correspondence principle \cite{Altschul:2007xf, Altschul:2007vca}.  So, by measuring the phase difference between the GW perturbations received at two detectors and associating this phase shift with the quantum phase shift of the graviton, we should be able to place bounds on the charge of the graviton using the above equation. Note that unlike the prescription described in the previous section, these bounds are free of any degeneracies with the mass of the graviton. \\

    In general, measuring 
    $\Delta \phi$ between two detectors for a particular GW signal is difficult. 
    For equal-mass, face-on GW signals that are well-described by the dominant $(\ell=m=2)$ harmonic of the GW signal, $\Delta \phi$ is perfectly degenerate with the reference phase at coalescence $\phi_c$, and any finite $\Delta \phi$ can be absorbed as a redefinition of $\phi_c$. However, when non-quadropolar modes are relevant for describing the GW signal (e.g. for unequal mass binaries or highly inclined binaries), the degeneracy between $\phi_c$ and $\Delta \phi$ is broken, and it is possible to measure $\Delta \phi$. This effect is already well-known in the literature in works dealing with strong gravitational lensing of GWs. Strong lensing naturally produces ``Type-II'' images that have a phase shift of $\pi/2$ relative to an un-lensed signal~\cite{Schneider:1992bmb, Dai:2017huk}, and this phase shift has been shown to be measurable in the presence of higher harmonics of the GW radiation, orbital precession and eccentricity in the GW signal~\cite{Ezquiaga:2020gdt, Wang:2021kzt, Vijaykumar:2022dlp, Janquart:2021nus}. Hence, by making appropriate choices of the GW source, we can bound $\Delta\phi$ and thus the graviton charge independent of the mass.  \\

    \section{Constraining the graviton charge}
    \label{sec:results}

    \subsection{Assumptions on $B$ and $\lambda_B$}
    
    It is apparent from Eq.~\eqref{effm} and Eq.~\eqref{eq:aharonov-bohm-constraint} that the bounds on $q_g$ will become tighter with increasing $B$ and $\lambda_B$.
    While the value of $B$ and $\lambda_B$ is not known, many observations have placed lower limits on these parameters (see~\cite{Neronov:2013zka, Subramanian:2015lua} for a review). For instance, Ref.~\cite{Neronov:2010gir}
    obtained a lower limit on $B$ exploiting the non-observation of GeV gamma-ray emission from the electromagnetic cascade initiated by TeV gamma-rays (from blazars) to be $B> 3 \times 10^{-16}\, {\rm G}$ while being coherent over Mpc scales~\cite{Neronov:2010gir}. More specifically, these constraints state,
    \begin{equation}
        B > \begin{cases}
          3 \times 10^{-16}\left(\dfrac{\lambda_B}{1\,{\rm Mpc}}\right)^{-1/2}\, {\rm G} \,,  & \lambda_B \leq 1\,{\rm Mpc}  \\
          3 \times 10^{-16}\, {\rm G} \,,  & \lambda_B > 1\,{\rm Mpc} 
        \end{cases}
    \end{equation}
    While there have been some improvements to these constraints over the years, we shall make a conservative choice $B = 3 \times 10^{-16}\,{\rm G}$ and $\lambda_B = 1\,{\rm Mpc}$ for our results.    

    \subsection{Results}
    
    \begin{figure}
    \label{fig:mg-qg-GWTC3}
        \includegraphics[width=0.5\textwidth]{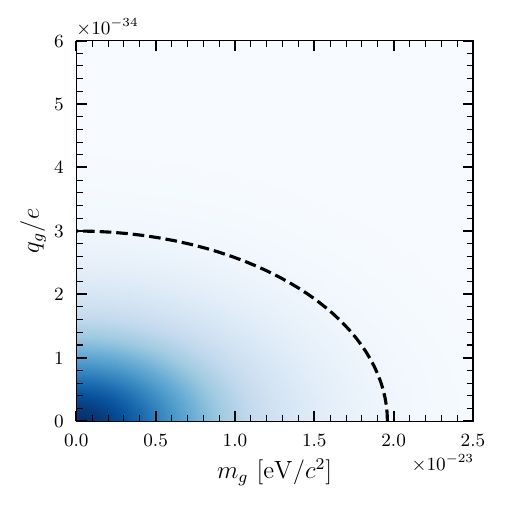}
        \includegraphics[width=0.5\textwidth]{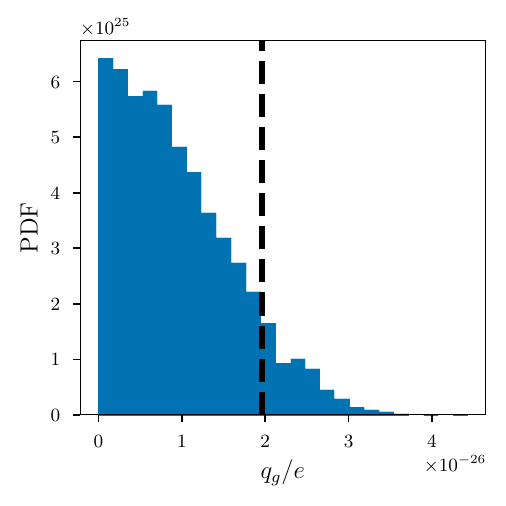}

        \caption{(Left) Joint posterior on $m_g$ and $q_g/e$ parameter space as inferred from GWTC-3 data. The colour corresponds to the probability density---darker colour signifies a higher probability density. Since the measurement of $q_g$ and $m_g$ are degenerate following Eq.~\eqref{phsemod}, we see that the isoprobability contours are ellipses in this two-dimensional $m_g$ --- $q_g/e$ plane. The black dashed line is the 90\% isoprobability contour. Marginalizing over $m_g$, we infer a 90\% upper limit of $|q_g|/e < \ChargeConstraintMonePN{}$. (Right) Posterior probability distribution on $q_g/e$ as estimated from the measured phase difference between LIGO-Hanford and LIGO-Livingston for GW190814. At 90\% CL, we place an upper limit of $\ChargeConstraintAH{}$ on $q_g/e$. Although weaker than the constraints obtained from reinterpreting the graviton mass constraints, these constraints are independent and are not degenerate with constraints on the graviton mass.}        
    \end{figure}
    
    We use the posterior on the mass of the (uncharged) graviton combined over GWTC-3 events by LVK collaboration~\cite{LIGOScientific:2021sio} and reinterpret it in the context of a charged massive graviton as discussed above. We report constraints on the joint $m_g$--- $q_g$ parameter space, with results shown in Fig.~\ref{fig:mg-qg-GWTC3}. As we see, the isoprobability contours in the image form ellipses, which follows directly from Eq.~\eqref{effm}. We also show the ellipse corresponding to the $90\%$ contour. Under the aforementioned assumptions on  $B$ and $\lambda_B$, the charge of the graviton is constrained to be
    \begin{equation}
    \abs{q_g}/e < \ChargeConstraintMonePN{} \qq{.}
    \end{equation} 
    For a different value of $B$ and $\lambda_B$, these bounds will scale trivially as $(B \lambda_B)^{-1}$. We should note that these bounds assume a flat prior on $m_{\rm eff}$. One could alternatively recalculate the bounds with individually flat priors on $m_g$ and $q_g$, which induces a prior $\pi(m_{\rm eff}) \propto m_{\rm eff}$. Since this prior is quite different from a flat prior, obtaining results with it would require rerunning the parameter inference on each event to effectively sample the parameter space. At most, we expect these results to change by an $\order{1}$ number, and refrain from carrying out this exercise.\\

We will next constrain the charge independently of the mass using Eq.~\eqref{eq:aharonov-bohm-constraint}. This method obviously has the advantage that we make no reference to the constraints on the graviton mass since the Aharonov-Bohm type phase shift can only occur in the presence of a non-zero charge. Even though this method is independent, the constraints are in general weaker than the one obtained using GW dephasing. As mentioned earlier, the measurement of phase difference between GW detectors can only be achieved with events that have significant higher harmonic content, failing which the phase difference measurement is exactly degenerate with the measurement of the coalescence phase of the binary $\phi_c$. We hence place bounds on $q_g$ using GW190814~\cite{LIGOScientific:2020zkf}, an asymmetric mass binary that showed significant evidence for higher-mode content in the signal. Although GW190814 was detected as a coincident signal by LIGO-Hanford, LIGO-Livingston, and Virgo detectors, we only use data from the LIGO detectors for our constraint due to the relatively low SNR in the Virgo detector\footnote{ While we have used only the phase difference between two detectors for our constraints with GW190814, this prescription can be easily generalized to multiple detectors by parametrizing all the relative phase differences through $q_g$. }.
We estimate the posterior on $\Delta \phi$ using the \texttt{bilby}~\cite{Ashton:2018jfp} software package, using the nested sampler \texttt{dynesty}~\cite{Speagle:2019ivv} to sample the posterior distribution. We use the waveform approximant IMRPhenomXPHM~\cite{Pratten:2020ceb} implemented in \texttt{lalsuite}~\cite{lalsuite}, and use the same standard priors as used in Ref.~\cite{LIGOScientific:2020zkf} along with a flat prior on $\Delta \phi$ in the range $[-\pi, \pi]$. We use the relative binning likelihood~\cite{relbin_cornish, relbin_zackay,Cornish:2021lje} implemented in \texttt{bilby}~\cite{Krishna:2023bug} to speed up the parameter inference\footnote{While the relative binning implementation in \texttt{bilby} is strictly optimal for signals dominated by the quadrupolar mode of GW radiation, we have verified that the likelihood errors introduced by the non-optimality are small for our purposes. This is also in agreement with results of Ref.~\cite{Krishna:2023bug}.}, and use \texttt{bilby\_pipe}~\cite{Romero-Shaw:2020owr} to streamline the inference runs. We obtain $\Delta \phi = -0.05^{+0.2}_{-0.25}$ at 90\% CL which we consequently use to place an bound 
\begin{equation}
\abs{q_g}/e < \ChargeConstraintAH{}
\end{equation}
at 90\% CL---eight orders of magnitude weaker than the constraints we obtained before. This bound is driven mainly by the $\order{0.1}$ uncertainty in the measurement of $\Delta \phi$.
    
    \section{Conclusion}
    \label{sec:summary}

    In this paper, we considered the possibility of a graviton endowed with a non-zero $U(1)$ charge. In the light of the Weinberg-Witten theorem, a non-zero $U(1)$ charge may make sense if the graviton is massive. A non-zero magnetic field will affect the trajectory of a charged graviton; we use this fact to compute the correction to the GW phasing that arises due to a non-zero charge. 
    Our analysis points at a degeneracy of the graviton mass bounds from \cite{LIGOScientific:2021sio} with the possible graviton charge. Using the LVK bounds on the mass of the graviton, we can impose an upper bound on the U(1) charge of a graviton in terms of the electron charge as  $|q_g|/e < \ChargeConstraintMonePN{}$.\\
    
    In addition to the above bound, we also use the total phase difference between the GW detections at LIGO-Hanford and LIGO-Livingston for the event GW190814 to find another independent upper bound of the graviton charge. This is done by identifying the phase difference with the possible Aharonov-Bohm phase shift experienced by a $U(1)$ charged graviton \cite{Altschul:2007vca, Altschul:2007xf}. This provides us a novel bound on the graviton charge independent of its mass as $|q_g|/e < \ChargeConstraintAH{}$.  Similar to the bound obtained on the charge of a photon, the derivation of this independent bound on the graviton charge assumes the quantum phase difference arising due to a non-zero charge coincides with the classical phase difference of the two separate GW detections through the correspondence principle.\\
    
    In the near future, the constraints on $q_g$ will improve thanks to the larger baseline $d$ afforded by the addition of LIGO-India~\cite{LIGO-INDIA, Saleem:2021iwi} to the GW detector network. These constraints will also improve by a few orders of magnitude with even larger baselines and/or sensitivities afforded by the next generation of ground-based (Cosmic Explorer~\cite{Reitze:2019iox}, Einstein Telescope~\cite{Punturo:2010zz}, etc.) and space-based (LISA~\cite{Danzmann:2003tv}, DECIGO~\cite{Sato:2017dkf}, etc.) detectors. Additionally, detections of more binaries with significant higher harmonics, eccentricity, and orbital precession will help in the measurement of the relative phase difference between detectors, hence also providing better constraints on $q_g$.  Our analysis uses conservative estimates for the intergalactic magnetic field strength and the correlation length given by Ref.~\cite{Neronov:2010gir}, and will improve once the lower bounds on these quantities becomes stricter, or if conclusive measurements of these quantities are made.

\appendix
\section{Calculation of phase at detector.} 
\label{Appendix:Phase at detector.}

    The total additional time that a charged massive graviton would take to propagate from the source to the detector can be calculated by plugging in $\delta t^m _{f_e,f_{e'}}$ and $\delta t^q _{f_e,f_{e'}}$ of Eq.\eqref{tm} and Eq.\eqref{tq} in Eq.\eqref{deltato} resulting in
    \begin{equation}
    \label{dto}
        \begin{aligned}
            \Delta t_o&\approx\Delta t_e+\frac{D_Lc^3m_g^2}{2h^2}\left(\frac{1}{f_e^2}\right)+\frac{2}{3}\frac{c D_L \lambda_B^2 B^2 q_g^2}{4 !h^2 }\left(\frac{1}{f_e^2}\right)-L(f_{e'})
        \end{aligned}
    \end{equation} 
    where,
    \begin{equation*}
        L(f_{e'})=\frac{D_Lc^3m_g^2}{2h^2}\left(\frac{1}{f_{e'}^2}\right)+\frac{2}{3}\frac{c D_L \lambda_B^2 B^2 q_g^2}{4 !h^2 }\left(\frac{1}{f_{e'}^2}\right)~.
    \end{equation*}
    Next, we may observe that the GW phasing in the frequency domain at the detector frame can be expressed as the following integral \cite{Blanchet:2001aw, Blanchet:2013haa}
    \begin{equation}
    \label{phase}
    \psi(f)=2 \pi \int_{f'}^{f}\left(t-t_c   \right) \, df + 2 \pi f_e t_c -\phi_c - \frac{\pi}{4}~.
    \end{equation}    
    The contribution of the charge and the mass of a graviton on the GW phase can be obtained by replacing $t-t_c$ with $\Delta t_o$ from Eq.~\eqref{dto} and carrying out the integral. Under the assumption that the effect of the propagation over the large distance $D_L$ will dominate the corrections due to the back reaction in the modified theory \cite{Will:1997bb, Ghosh:2023xes}, we will get
    \begin{equation}
        \begin{aligned}
            \psi(f)= & 2 \pi f \tilde{t}_c-\tilde{\Phi}_c-\frac{\pi}{4} + \sum_n\alpha^n f^\frac{2n}{3}-\left\{\frac{\pi  m_g^2 c^3}{h^2}+\left(\frac{2}{3}B^2\lambda_B^2\right)\frac{2 \pi q_g^2c }{4 ! h^2}\right\}\frac{D_L}{f}.\\
        \end{aligned}
    \end{equation}      


\acknowledgments
    We thank R Loganayagam and Alok Laddha for discussions regarding field theoretic aspects of a charged graviton. We are grateful to James Beattie and Kandaswamy Subramanian for their valuable insights into large-scale magnetic fields. We also appreciate helpful comments by Brett Altschul and Haris M K.
    
    The Research of SN is supported by the Prime Minister's Research Fellowship (ID-1701653), Government of India. AV is supported by the Natural Sciences and Engineering Research Council of Canada (NSERC) (funding reference number 568580). The research of SS is supported by the Department of Science and Technology, Government of India, under the SERB CRG Grant (No. CRG/2023/000934). AV would like to thank IIT Gandhinagar for hospitality during his stay when this work was initiated. SS conveys his gratitude to Instituto Galego de F´ısica de Altas Enerx´ıas (IGFAE), Spain for wonderful hospitality during his stay on a sabbatical leave. 

    This material is based upon work supported by NSF's LIGO Laboratory which is a major facility fully funded by the National Science Foundation. We acknowledge use of numpy~\citep{2020Natur.585..357H}, scipy~\citep{2020NatMe..17..261V}, matplotlib~\citep{2007CSE.....9...90H}, astropy~\citep{2013AA...558A..33A, 2018AJ....156..123A}, jupyter~\citep{2016ppap.book...87K}, pandas~\citep{mckinney-proc-scipy-2010}, seaborn~\citep{2021JOSS....6.3021W}, {bilby}~\cite{Ashton:2018jfp}, {bilby\_pipe}~\cite{Romero-Shaw:2020owr}, 
 {dynesty}~\cite{Speagle:2019ivv}, {pesummary}~\cite{Hoy:2020vys}, and lalsuite~\citep{lalsuite} software packages.
    \\




\begin{thebibliography}{99}

\bibitem{LIGOScientific:2016aoc}
B.~P.~Abbott \textit{et al.} [LIGO Scientific and Virgo],
\emph{Observation of Gravitational Waves from a Binary Black Hole Merger,
Phys. Rev. Lett.} \textbf{116}, no.6, 061102 (2016),
[arXiv:1602.03837 [gr-qc]].

\bibitem{LIGOScientific:2018mvr}
B.~P.~Abbott \textit{et al.} [LIGO Scientific and Virgo],
\emph{GWTC-1: A Gravitational-Wave Transient Catalog of Compact Binary Mergers Observed by LIGO and Virgo during the First and Second Observing Runs,
Phys. Rev. X} \textbf{9}, no.3, 031040 (2019),
[arXiv:1811.12907 [astro-ph.HE]].

\bibitem{LIGOScientific:2020ibl}
R.~Abbott \textit{et al.} [LIGO Scientific and Virgo],
\emph{GWTC-2: Compact Binary Coalescences Observed by LIGO and Virgo During the First Half of the Third Observing Run,
Phys. Rev. X} \textbf{11}, 021053 (2021),
[arXiv:2010.14527 [gr-qc]].

\bibitem{KAGRA:2021vkt}
R.~Abbott \textit{et al.} [KAGRA, VIRGO and LIGO Scientific],
\emph{GWTC-3: Compact Binary Coalescences Observed by LIGO and Virgo during the Second Part of the Third Observing Run,
Phys. Rev. X} \textbf{13}, no.4, 041039 (2023),
[arXiv:2111.03606 [gr-qc]].

\bibitem{LIGOScientific:2014pky}
J.~Aasi \textit{et al.} [LIGO Scientific],
\emph{Advanced LIGO,
Class. Quant. Grav.} \textbf{32}, 074001 (2015),
[arXiv:1411.4547 [gr-qc]].

\bibitem{VIRGO:2014yos}
F.~Acernese \textit{et al.} [VIRGO],
\emph{Advanced Virgo: a second-generation interferometric gravitational wave detector,
Class. Quant. Grav.} \textbf{32}, no.2, 024001 (2015),
[arXiv:1408.3978 [gr-qc]].

\bibitem{KAGRA:2020tym}
T.~Akutsu \textit{et al.} [KAGRA],
\emph{Overview of KAGRA: Detector design and construction history,
PTEP} \textbf{2021}, no.5, 05A101 (2021),
[arXiv:2005.05574 [physics.ins-det]].

\bibitem{LIGOScientific:2016lio}
B.~P.~Abbott \textit{et al.} [LIGO Scientific and Virgo],
\emph{Tests of general relativity with GW150914,
Phys. Rev. Lett.} \textbf{116}, no.22, 221101 (2016)
[erratum: Phys. Rev. Lett. \textbf{121}, no.12, 129902 (2018)],
[arXiv:1602.03841 [gr-qc]].

\bibitem{LIGOScientific:2018dkp}
B.~P.~Abbott \textit{et al.} [LIGO Scientific and Virgo],
\emph{Tests of General Relativity with GW170817,
Phys. Rev. Lett.} \textbf{123}, no.1, 011102 (2019),
[arXiv:1811.00364 [gr-qc]].

\bibitem{LIGOScientific:2019fpa}
B.~P.~Abbott \textit{et al.} [LIGO Scientific and Virgo],
\emph{Tests of General Relativity with the Binary Black Hole Signals from the LIGO-Virgo Catalog GWTC-1,
Phys. Rev. D} \textbf{100}, no.10, 104036 (2019),
[arXiv:1903.04467 [gr-qc]].

\bibitem{LIGOScientific:2020tif}
R.~Abbott \textit{et al.} [LIGO Scientific and Virgo],
\emph{Tests of general relativity with binary black holes from the second LIGO-Virgo gravitational-wave transient catalog,
Phys. Rev. D} \textbf{103}, no.12, 122002 (2021),
[arXiv:2010.14529 [gr-qc]].

\bibitem{LIGOScientific:2021sio}
R.~Abbott \textit{et al.} [LIGO Scientific, VIRGO and KAGRA],
\emph{Tests of General Relativity with GWTC-3,}
[arXiv:2112.06861 [gr-qc]].

\bibitem{Will:1997bb}
C.~M.~Will,
\emph{Bounding the mass of the graviton using gravitational wave observations of inspiralling compact binaries,
Phys. Rev. D} \textbf{57}, 2061-2068 (1998),
[arXiv:gr-qc/9709011 [gr-qc]].

\bibitem{Mirshekari:2011yq}
S.~Mirshekari, N.~Yunes and C.~M.~Will,
\emph{Constraining Generic Lorentz Violation and the Speed of the Graviton with Gravitational Waves,
Phys. Rev. D} \textbf{85}, 024041 (2012)
[arXiv:1110.2720 [gr-qc]].

\bibitem{Barbiellini:1987zz}
G.~Barbiellini and G.~Cocconi,
\emph{Electric Charge of the Neutrinos from SN1987A,
Nature} \textbf{329}, 21-22 (1987)

\bibitem{Raffelt:1994rb}
G.~Raffelt,
\emph{Pulsar bound on the photon electric charge reexamined,
Phys. Rev. D} \textbf{50}, 7729-7730 (1994),
[arXiv:hep-ph/9409461 [hep-ph]].

\bibitem{Altschul:2007xf}
B.~Altschul,
\emph{Bound on the Photon Charge from the Phase Coherence of Extragalactic Radiation,
Phys. Rev. Lett.} \textbf{98}, 261801 (2007)
[arXiv:hep-ph/0703126 [hep-ph]].

\bibitem{Altschul:2007vca}
B.~Altschul,
\emph{Astrophysical Bounds on the Photon Charge and Magnetic Moment,
Astropart. Phys.} \textbf{29}, 290-298 (2008),
[arXiv:0711.2038 [hep-th]].

\bibitem{Cocconi:1988xi}
G.~Cocconi,
\emph{Upper Limit for the Electric Charge of the Photons From the Millisecond Pulsar 1937+21 Observations,
Phys. Lett. B} \textbf{206}, 705-706 (1988),

\bibitem{Berends:1979rv}
F.~A.~Berends, J.~W.~van Holten, P.~van Nieuwenhuizen and B.~de Wit,
\emph{``On Field Theory for Massive and Massless Spin 5/2 Particles,''
Nucl. Phys. B} \textbf{154}, 261-282 (1979)

\bibitem{deWit:1979sib}
B.~de Wit and D.~Z.~Freedman,
\emph{Systematics of Higher Spin Gauge Fields,
Phys. Rev. D} \textbf{21}, 358 (1980)

\bibitem{Johnson:1960vt}
K.~Johnson and E.~C.~G.~Sudarshan,
\emph{Inconsistency of the local field theory of charged spin 3/2 particles,
Annals Phys.} \textbf{13}, 126-145 (1961)

\bibitem{Weinberg:1980kq}
S.~Weinberg and E.~Witten,
\emph{Limits on Massless Particles,
Phys. Lett. B} \textbf{96}, 59-62 (1980)

\bibitem{Aharonov:1959fk}
Y. Aharonov and D. Bohm, \emph{Significance of electromagnetic potentials in the quantum theory, Phys. Rev.} \textbf{115}, 485–491 (1959).

\bibitem{Blanchet:2001aw}
L.~Blanchet, B.~R.~Iyer and B.~Joguet,
\emph{Gravitational waves from inspiralling compact binaries: Energy flux to third postNewtonian order,
Phys. Rev. D }\textbf{65}, 064005 (2002)
[erratum: Phys. Rev. D \textbf{71}, 129903 (2005)],
[arXiv:gr-qc/0105098 [gr-qc]].

\bibitem{Blanchet:2013haa}
L.~Blanchet,
\emph{Gravitational Radiation from Post-Newtonian Sources and Inspiralling Compact Binaries,
Living Rev. Rel.} \textbf{17}, 2 (2014),
[arXiv:1310.1528 [gr-qc]].

\bibitem{Ghosh:2023xes}
R.~Ghosh, S.~Nair, L.~Pathak, S.~Sarkar and A.~S.~Sengupta,
\emph{Does the speed of gravitational waves depend on the source velocity?,
Phys. Rev. D} \textbf{108}, no.12, 124017 (2023),
[arXiv:2304.14820 [gr-qc]].

\bibitem{Schneider:1992bmb}
P.~Schneider, J.~Ehlers and E.~E.~Falco,
\emph{Gravitational Lenses,
Springer,} (1992).

\bibitem{Dai:2017huk}
L.~Dai and T.~Venumadhav,
\emph{On the waveforms of gravitationally lensed gravitational waves},
[arXiv:1702.04724 [gr-qc]].

\bibitem{Ezquiaga:2020gdt}
J.~M.~Ezquiaga, D.~E.~Holz, W.~Hu, M.~Lagos and R.~M.~Wald,
\emph{Phase effects from strong gravitational lensing of gravitational waves,
Phys. Rev. D }\textbf{103}, no.6, 064047 (2021),
[arXiv:2008.12814 [gr-qc]].

\bibitem{Wang:2021kzt}
Y.~Wang, R.~K.~L.~Lo, A.~K.~Y.~Li and Y.~Chen,
\emph{Identifying Type II Strongly Lensed Gravitational-Wave Images in Third-Generation Gravitational-Wave Detectors,
Phys. Rev. D }\textbf{103}, no.10, 104055 (2021),
[arXiv:2101.08264 [gr-qc]].

\bibitem{Vijaykumar:2022dlp}
A.~Vijaykumar, A.~K.~Mehta and A.~Ganguly,
\emph{Detection and parameter estimation challenges of type-II lensed binary black hole signals,
Phys. Rev. D }\textbf{108}, no.4, 043036 (2023),
[arXiv:2202.06334 [gr-qc]].

\bibitem{Janquart:2021nus}
J.~Janquart, E.~Seo, O.~A.~Hannuksela, T.~G.~F.~Li and C.~V.~Broeck,
\emph{On the Identification of Individual Gravitational-wave Image Types of a Lensed System Using Higher-order Modes,
Astrophys. J. Lett.} \textbf{923}, no.1, L1 (2021),
[arXiv:2110.06873 [gr-qc]].

\bibitem{Neronov:2013zka}
A.~Neronov, A.~M.~Taylor, C.~Tchernin and I.~Vovk,
\emph{Measuring the correlation length of intergalactic magnetic fields from observations of gamma-ray induced cascades,
Astron. Astrophys.} \textbf{554}, A31 (2013),
[arXiv:1307.2753 [astro-ph.HE]].

\bibitem{Subramanian:2015lua}
K.~Subramanian,
\emph{The origin, evolution and signatures of primordial magnetic fields,
Rept. Prog. Phys. }\textbf{79}, no.7, 076901 (2016),
[arXiv:1504.02311 [astro-ph.CO]].

\bibitem{Neronov:2010gir}
A.~Neronov and I.~Vovk,
\emph{Evidence for strong extragalactic magnetic fields from Fermi observations of TeV blazars,
Science} \textbf{328}, 73-75 (2010),
[arXiv:1006.3504 [astro-ph.HE]].

\bibitem{LIGOScientific:2020zkf}
R.~Abbott \textit{et al.} [LIGO Scientific and Virgo],
\emph{GW190814: Gravitational Waves from the Coalescence of a 23 Solar Mass Black Hole with a 2.6 Solar Mass Compact Object,
Astrophys. J. Lett. }\textbf{896}, no.2, L44 (2020),
[arXiv:2006.12611 [astro-ph.HE]].

\bibitem{Ashton:2018jfp}
G.~Ashton, M.~H\"ubner, P.~D.~Lasky, C.~Talbot, K.~Ackley, S.~Biscoveanu, Q.~Chu, A.~Divakarla, P.~J.~Easter and B.~Goncharov, \textit{et al.}
\emph{BILBY: A user-friendly Bayesian inference library for gravitational-wave astronomy,
Astrophys. J. Suppl.} \textbf{241}, no.2, 27 (2019),
[arXiv:1811.02042 [astro-ph.IM]].

\bibitem{Speagle:2019ivv}
Speagle, Joshua S. \emph{dynesty: a dynamic nested sampling package for estimating Bayesian posteriors and evidences. Monthly Notices of the Royal Astronomical Society} \textbf{493}, no. 3, 3132-3158 (2020), arXiv:1904.02180 [astro-ph.IM].

\bibitem{Pratten:2020ceb}
G.~Pratten, C.~Garc\'\i{}a-Quir\'os, M.~Colleoni, A.~Ramos-Buades, H.~Estell\'es, M.~Mateu-Lucena, R.~Jaume, M.~Haney, D.~Keitel and J.~E.~Thompson, \textit{et al.}
\emph{Computationally efficient models for the dominant and subdominant harmonic modes of precessing binary black holes,
Phys. Rev. D} \textbf{103}, no.10, 104056 (2021),
[arXiv:2004.06503 [gr-qc]].

\bibitem{lalsuite}
LIGO Scientific Collaboration, Virgo Collaboration, and
KAGRA Collaboration, \emph{LVK Algorithm Library - LAL-
Suite, Free software (GPL)} (2018).

\bibitem{relbin_cornish}
N.~J.~Cornish,
\emph{Fast Fisher Matrices and Lazy Likelihoods,} [arXiv:1007.4820 [gr-qc]].

\bibitem{relbin_zackay}
B.~Zackay, L.~Dai and T.~Venumadhav,
\emph{Relative Binning and Fast Likelihood Evaluation for Gravitational Wave Parameter Estimation},
[arXiv:1806.08792 [astro-ph.IM]].

\bibitem{Cornish:2021lje}
N.~J.~Cornish,
\emph{Heterodyned likelihood for rapid gravitational wave parameter inference,
Phys. Rev. D }\textbf{104}, no.10, 104054 (2021),
[arXiv:2109.02728 [gr-qc]].

\bibitem{Krishna:2023bug}
K.~Krishna, A.~Vijaykumar, A.~Ganguly, C.~Talbot, S.~Biscoveanu, R.~N.~George, N.~Williams and A.~Zimmerman,
\emph{Accelerated parameter estimation in Bilby with relative binning, }
[arXiv:2312.06009 [gr-qc]].

\bibitem{Romero-Shaw:2020owr}
I.~M.~Romero-Shaw, \textit{et al.}
\emph{Bayesian inference for compact binary coalescences with bilby: validation and application to the first LIGO\textendash{}Virgo gravitational-wave transient catalogue,
Mon. Not. Roy. Astron. Soc.} \textbf{499}, no.3, 3295-3319 (2020),
[arXiv:2006.00714 [astro-ph.IM]].

\bibitem{LIGO-INDIA}
C. S. Unnikrishnan,\emph{ Indigo and ligo-india: Scope and
plans for gravitational wave research and precision metrology in india, International Journal of Modern Physics D }\textbf{22}, 1341010 (2013).

\bibitem{Saleem:2021iwi}
M.~Saleem, \textit{et al.}
\emph{The science case for LIGO-India,
Class. Quant. Grav.} \textbf{39}, no.2, 025004 (2022),
[arXiv:2105.01716 [gr-qc]].

\bibitem{Reitze:2019iox}
D.~Reitze, \textit{et al.}
\emph{Cosmic Explorer: The U.S. Contribution to Gravitational-Wave Astronomy beyond LIGO,
Bull. Am. Astron. Soc. }\textbf{51}, no.7, 035 (2019),
[arXiv:1907.04833 [astro-ph.IM]].

\bibitem{Punturo:2010zz}
M. Punturo \textit{et al.}, \emph{The Einstein Telescope: A third-
generation gravitational wave observatory, Class. Quant.
Grav. }\textbf{27}, 194002 (2010).

\bibitem{Danzmann:2003tv}
K. Danzmann and A. Rudiger, \emph{LISA technology - Concept, status, prospects, Class. Quant. Grav.} \textbf{20}, S1–S9 (2003).

\bibitem{Sato:2017dkf}
Shuichi Sato et al., \emph{The status of DECIGO, J. Phys. Conf. Ser.} \textbf{840}, 012010 (2017).

\bibitem{2020Natur.585..357H}
C.~R.~Harris, \textit{et al.}
\emph{Array programming with NumPy,
Nature} \textbf{585}, no.7825, 357-362 (2020),
[arXiv:2006.10256 [cs.MS]].

\bibitem{2020NatMe..17..261V}
Pauli Virtanen \textit{et al.}, \emph{SciPy 1.0–Fundamental Algorithms for Scientific Computing in Python Nature Meth.} \textbf{17}, 261 (2020), arXiv:1907.10121 [cs.MS].

\bibitem{2007CSE.....9...90H}
John D. Hunter, \emph{Matplotlib: A 2D Graphics Environment, Computing in Science and Engineering} \textbf{9}, 90–95
(2007).

\bibitem{2013AA...558A..33A}
Thomas P. Robitaille \textit{et al.} \emph{(Astropy), Astropy: A Community Python Package for Astronomy,” Astron. Astrophys.} \textbf{558}, A33 (2013), arXiv:1307.6212 [astro-ph.IM].

\bibitem{2018AJ....156..123A}
A. M. Price-Whelan \textit{et al.} \emph{(Astropy), “The Astropy Project: Building an Open-science Project and Status of the v2.0 Core Package,” Astron. J.} \textbf{156}, 123 (2018),
arXiv:1801.02634.

\bibitem{2016ppap.book...87K}
Thomas Kluyver \textit{et. al.}, \emph{Jupyter Notebooks—a publishing format for reproducible computational work- flows,  IOS Press }, pp. 87–90, (2016).

\bibitem{mckinney-proc-scipy-2010}
Wes McKinney, \emph{Data Structures for Statistical Com- puting in Python, in Proceedings of the 9th Python in Science Conference, edited by Stéfan van der Walt and Jarrod Millman},pp. 56 – 61 (2010).

\bibitem{2021JOSS....6.3021W}
Michael Waskom, \emph{seaborn: statistical data visualization, The Journal of Open Source Software} \textbf{6}, 3021 (2021).

\bibitem{Hoy:2020vys}
Charlie Hoy and Vivien Raymond, \emph{PESummary: the code agnostic Parameter Estimation Summary page builder, SoftwareX} \textbf{15}, 100765 (2021), arXiv:2006.06639 [astro-ph.IM].

\end{thebibliography}



\end{document}